\documentclass[showpacs,aps,12pt]{revtex4}
\usepackage{bbm}%
\usepackage{amsfonts}%
\usepackage{amsmath}%
\usepackage{graphicx}%
\usepackage{amssymb}%

\begin{document}
\title{Near-threshold $\eta$ production in $pp$ collisions }

\author{Qi-Fang L\"{u} }
\affiliation{Department of Physics, Zhengzhou University, Zhengzhou,
Henan 450001, China}

\author{De-Min Li}\email{lidm@zzu.edu.cn}
\affiliation{Department of Physics, Zhengzhou University, Zhengzhou,
Henan 450001, China}

\begin{abstract}

We study near-threshold $\eta$ meson production in $pp$
collisions within an effective Lagrangian approach combined with the isobar
model, by allowing for the various intermediate nucleon resonances
due to the $\pi$, $\eta$, and $\rho$-meson exchanges. It is shown
that the $\rho$-meson exchange is the dominant excitation mechanism
for these resonances, and the contribution from the $N^*(1720)$ is
dominant. The total cross section data can be reasonably reproduced, and the
anisotropic angular distributions of the emitted $\eta$ meson are
consistent with experimental measurements. Besides, the invariant
mass spectra of $pp$ and $p\eta$ explain the data well at excess
energy of 15 MeV, and are basically consistent with the data at
excess energy of 40 MeV. However, our model calculations cannot
reasonably account for the two-peak structure in the $p\eta$
distribution at excess energies of 57 and 72 MeV, which suggests that a more
complicated mechanism is needed at higher energy region.

\end{abstract}

\maketitle

\section{Introduction}{\label{introduction}}

The properties of hadrons are subject to the behavior of quantum
chromodynamics (QCD) in the non-perturbative region, so the study of
hadrons is important to deepen the understanding of the
non-perturbative properties of QCD. Meson production reactions in
nucleon-nucleon collisions near threshold are a good platform to
obtain new information on the hadrons and therefore have received a
lot of attentions both experimentally and theoretically in recent
years~\cite{hanhart}. A large set of experimental data on $\eta$
meson production in the $pp\to
pp\eta$~\cite{eta_exp1,eta_exp2,eta_exp3,eta_exp4,eta_exp5,eta_exp6,petr,Rafal,
Pauly}, $pn\to d\eta$, and $pn\to pn\eta$
reactions~\cite{Calen1,Calen2,Calen3,moskal2009} has been
accumulated, which has stimulated many theoretical investigations on
$\eta$ meson production
~\cite{bati,geda,germond,laget,vetter,alvaredo,pena,shyam07,
gedalin,san,faldt,baru,nakayama,knakayama,bernard,moskal02,riska,xucao,Nakayama:2008tg,deloffprc69}.

$\eta$ meson production in $NN$ collisions is generally
assumed to occur predominantly through re-scattering of the
intermediate nucleon resonances caused by the meson exchanges. For
this basic mechanism, it is not yet clear which of the possible
nucleon resonances plays the dominant role. For example, in
Refs.~\cite{bati,geda,germond,laget,vetter,shyam07,gedalin,san,faldt,baru,nakayama,knakayama}
it is suggested that the $N^*(1535)$ is dominant, while in Ref.~\cite{Nakayama:2008tg} it is found that the $N^*(1520)$ is dominant
and the $N^*(1535)$ contribution is small due to the strong
destructive interference among the exchanged mesons. Also, it is not
yet clear which of the possible meson exchanges plays the most
important role. For example, in Refs.~\cite{bati,shyam07,xucao} it is
claimed that the pseudoscalar mesons $\pi$ and $\eta$ exchanges are
dominant, whereas in Refs.~\cite{geda,san,faldt,vetter} it is
suggested that vector meson $\rho$ exchange is dominant. In
Ref.~\cite{nakayama} it is found that both the $\pi$ exchange and
the $\rho$ exchange can describe the cross sections well. The study
on the $N^*(1535)N\rho$ coupling in the framework of an effective
Lagrangian approach manifests that the value of $g_{N^*(1535)N\rho}$
is strong~\cite{xierho}, which favors the importance of the $\rho$ meson exchange
in $NN$ collisions.

In Ref.~\cite{deloffprc69}, the final state interaction (FSI)
enhancement factor is considered and it is found that the measured
$pp$ and $\eta p$ effective mass spectra can be well reproduced by
allowing for a linear energy dependence in the leading $^3P_0$ $\to$
$^1S_0,s$ partial wave amplitude.

In Ref.~\cite{petr} it is suggested that the higher partial waves may
be important even at 15.5 MeV. Besides the $N^*(1535)$, the $\rho$
meson may also couple strongly to other higher resonances. The large
branching ratio and the small phase space for the
$N^*(1720)\rightarrow N\rho$ also suggest that the $N^*(1720)N\rho$
coupling is strong.

 With the inspiration of the factors mentioned
above, we shall restudy the $pp\rightarrow pp\eta$ reaction in an
effective Lagrangian approach combined with the isobar model. The
combination of the effective Lagrangian approach and the isobar
model turns out to be a good method to study hadron resonance
production in the $\pi N$, $NN$, and $\bar{K}N$
scattering~\cite{shyam07,xucao,Nakayama:2008tg,xierho,xiephi1,xiephi2,lv1,lv2,lv3}.
In the present work, we assume that the near threshold $\eta$ meson
production in proton-proton collisions is through the intermediate
$N^*(1535)$, $N^*(1650)$, $N^*(1710)$, $N^*(1720)$, and the nucleon
pole caused by the $\pi$, $\eta$, and $\rho$ mesons exchanges. The
proton-proton FSI and proton-$\eta$ FSI are also considered.

This work is organized as follows. The basic formalism and
ingredients used in our model are given in Section II. The numerical
results and discussions are given in Section III. A summary is
given in Section IV.

\section{Formalism and ingredients}{\label{formalism}}

\begin{figure}[htbp]
\includegraphics[scale=0.7]{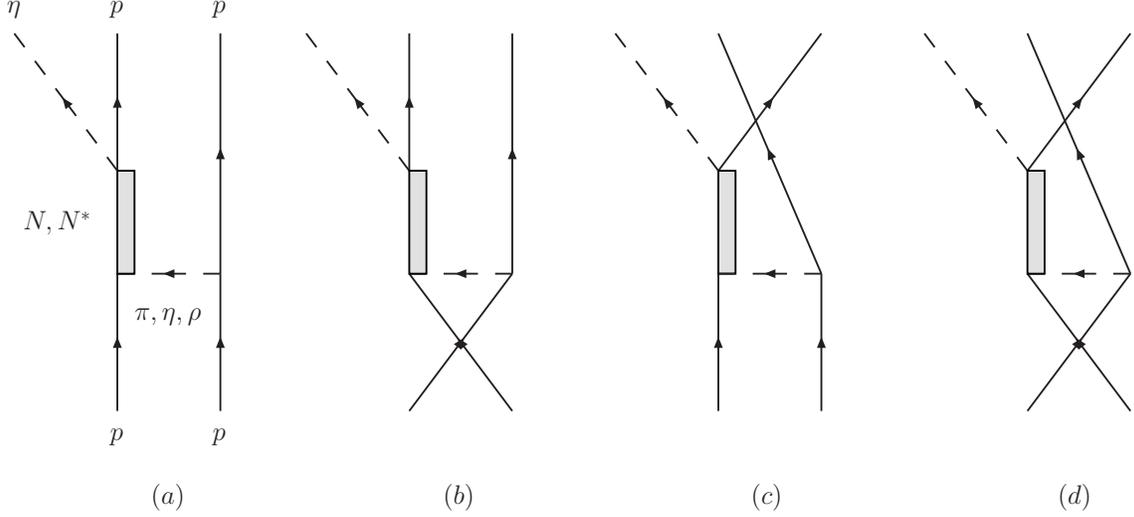}
\vspace{-0.2cm} \caption{Feynman diagrams for the $pp \to pp \eta$
reaction.} \label{diagram}
\end{figure}

The basic tree level Feynman diagrams for the $pp \to pp \eta$
reaction are shown in Fig.~\ref{diagram}.

The interaction Lagrangians for the $\pi NN$, $\eta NN$ and $\rho
NN$ couplings can be written as~\cite{xiephi1,xiephi2}:
\begin{equation}
{\cal L}_{\pi N N}  = -i g_{\pi N N} \bar{\psi}_N \gamma_5 \vec\tau
\cdot \vec\pi \psi_N, \label{pin}
\end{equation}
\begin{equation}
{\cal L}_{\eta N N}  = -i g_{\eta N N} \bar{\psi}_N \gamma_5 \eta
\psi_N, \label{etan}
\end{equation}
\begin{equation}
{\cal L}_{\rho N N} = -g_{\rho N N}
\bar{\psi}_N\left(\gamma_{\mu}+\frac{\kappa}{2m_N} \sigma_{\mu \nu}
\partial^{\nu}\right)\vec\tau \cdot \vec\rho^{\mu} \psi_N. \label{rhon}
\end{equation}

The effects of the non-point-like structures of exchanged mesons are taken
into account by introducing the following off-shell form factors
in the $MNN$ vertexes\cite{mach1,mach2,mach3}
\begin{equation}
F^{NN}_M(k^2_M)=\left(\frac{\Lambda^2_M-m_M^2}{\Lambda^2_M-
k_M^2}\right)^n,
\end{equation}
where $M$ denotes the exchanged meson; $n=1$ for $\pi^0 NN$ and
$\eta NN$ vertexes, $n=2$ for the $\rho^0 NN$ vertex; $k_M$, $m_M$,
and $\Lambda_M$ are the 4-momentum, mass, and cut-off parameter, respectively, for
the exchanged-meson $M$. The relevant parameters used
in our calculations are: $g_{\pi NN}^2/4\pi$ = 14.4, $g_{\eta
NN}^2/4\pi$ = 0.4, $g_{\rho NN}^2/4\pi$ = 0.9, $\kappa$ =
6.1~\cite{mach1,mach2,mach3,tsushima1,tsushima2,tsushima3,sibi1,sibi2,xiephi1,xiephi2},
$\Lambda_{\rho}$ = 1.85 GeV~\cite{xiephi1,xiephi2}, and
$\Lambda_{\pi}$ = $\Lambda_{\eta}$  = 0.8 GeV.

The following interaction Lagrangians involving the nucleon
resonances $N^*$ can be obtained within a Lorentz covariant orbital-spin
(L-S) scheme for the $N^* N M$ couplings~\cite{zouprc03}:
\begin{equation}
{\cal L}_{\pi N N^*(1535)}  =  ig_{\pi N N^*(1535)}\bar{\psi}_N \vec\tau
\cdot \vec\pi \psi_{N^*(1535)}+h.c.,\label{pin1535}
\end{equation}
\begin{equation}
 {\cal L}_{\eta N N^*(1535)} = ig_{\eta N N^*(1535)}\bar{\psi}_N \eta \psi_{N^*(1535)}+h.c.,
 \label{etan1535}
\end{equation}
\begin{equation}
 {\cal L}_{\rho N N^*(1535)} = ig_{\rho N N^*(1535)}\bar{\psi}_N \gamma_5 \left(\gamma_{\mu}-\frac{q_{\mu} \not \! q}{q^2}\right) \vec\tau
\cdot \vec\rho^{\mu}(p_{\rho}) \psi_{N^*(1535)}
+h.c.,\label{rhon1535}
\end{equation}
\begin{equation}
{\cal L}_{\pi N N^*(1650)}  =  ig_{\pi N N^*(1650)}\bar{\psi}_N \vec\tau
\cdot \vec\pi \psi_{N^*(1650)}+h.c.,\label{pin1650}
\end{equation}
\begin{equation}
 {\cal L}_{\eta N N^*(1650)} = ig_{\eta N N^*(1650)}\bar{\psi}_N \eta \psi_{N^*(1650)}+h.c.,
 \label{etan1650}
\end{equation}
\begin{equation}
 {\cal L}_{\rho N N^*(1650)} = ig_{\rho N N^*(1650)}\bar{\psi}_N \gamma_5 \left(\gamma_{\mu}-\frac{q_{\mu} \not \! q}{q^2}\right) \vec\tau
\cdot \vec\rho^{\mu}(p_{\rho}) \psi_{N^*(1650)}
+h.c.,\label{rhon1650}
\end{equation}
\begin{equation}
{\cal L}_{\pi N N^*(1710)}  = -i g_{\pi N N^*(1710)} \bar{\psi}_N \gamma_5 \vec\tau
\cdot \vec\pi \psi_{N^*(1710)}+h.c., \label{pin1710}
\end{equation}
\begin{equation}
{\cal L}_{\eta N N^*(1710)}  = -i g_{\eta N N^*(1710)} \bar{\psi}_N \gamma_5 \eta
\psi_{N^*(1710)}+h.c., \label{etan1710}
\end{equation}
\begin{equation}
{\cal L}_{\rho N N^*(1710)} = -g_{\rho N N^*(1710)}
\bar{\psi}_N\left(\gamma_{\mu}+\frac{\kappa}{2m_N} \sigma_{\mu \nu}
\partial^{\nu}\right)\vec\tau \cdot \vec\rho^{\mu} \psi_{N^*(1710)}+h.c., \label{rhon1710}
\end{equation}
\begin{equation}
{\cal L}_{\pi N N^{*}(1720)} = g_{\pi N N^*(1720)} \bar{\psi}_N \vec\tau
\cdot \partial^{\mu} {\vec\pi} \psi_{N^*(1720)\mu} + h.c., \label{pin1720}
\end{equation}
\begin{equation}
{\cal L}_{\eta N N^{*}(1720)} = g_{\eta N N^*(1720)} \bar{\psi}_N \partial^{\mu} \eta \psi_{N^*(1720)\mu} + h.c., \label{etan1720}
\end{equation}
\begin{equation}
{\cal L}_{\rho N N^{*}(1720)} = g_{\rho N N^*(1720)} \bar{\psi}_N \gamma_5 \vec\tau
\cdot \vec\rho^{\mu} \psi_{N^*(1720)\mu} + h.c.. \label{rhon1720}
\end{equation}

At the $N^*NM$ vertexes, the monopole form factors are employed:
\begin{equation}
F^{N^* N}_M(k^2_M)=\frac{\Lambda^{*2}_M-m_M^2}{\Lambda^{*2}_M-
k_M^2}. \label{sff}
\end{equation}

With the effective Lagrangians listed above, the partial width of
the nucleon resonance $N^*$ can be derived. From the experimental
data on the partial width of the corresponding nucleon resonance,
the $N^*NM$ coupling constants can be obtained. For the $N^*$
resonance below the $N\rho$ threshold, the partial decay width
$\Gamma_{N^* \to N \rho \to N \pi\pi }$ is employed to determine the
$N^*N\rho$ coupling constant~\cite{xiephi1,xiephi2}. The relevant
coupling constants and cut-off parameters are listed in
Table~\ref{tab1}.

\begin{table}
\begin{center}
\caption{ \label{tab1} Relevant parameters of the nucleon resonances used in our calculation.
 The widths and branching ratios are taken from the PDG~\cite{pdg2012}.}
\footnotesize
\begin{tabular}{|cccccc|}
\hline
 Resonance &  Width(GeV) & Decay channel  & Branching ratios  & $g^{2}/4\pi$ & Cut-off(GeV)\\
\hline
$N^*(1535)$ & 0.15 &$N\pi$\hphantom{00}  & \hphantom{0}0.45 & 0.037 & 0.8\\
 & &$N\eta$\hphantom{00}  & \hphantom{0}0.42 &0.28  & 0.8\\
 & &$N\rho$\hphantom{00}  & \hphantom{0}0.02 &5.55  & 0.8\\
$N^*(1650)$ & 0.15 &$N\pi$\hphantom{00}  & \hphantom{0}0.70 & 0.052 & 1.5\\
 & &$N\eta$\hphantom{00}  & \hphantom{0}0.10 &0.036  & 1.5\\
 & &$N\rho$\hphantom{00}  & \hphantom{0}0.01 &0.0064  & 1.5\\
$N^*(1710)$ & 0.1 &$N\pi$\hphantom{00}  & \hphantom{0}0.125 & 0.072 & 1.5\\
 & &$N\eta$\hphantom{00}  & \hphantom{0}0.20 &0.97  & 1.5\\
 & &$N\rho$\hphantom{00}  & \hphantom{0}0.15 &0.019  & 1.5\\
$N^*(1720)$ & 0.25 &$N\pi$\hphantom{00}  & \hphantom{0}0.11 & 0.11 & 1.5\\
 & &$N\eta$\hphantom{00}  & \hphantom{0}0.04 &0.35  & 1.5\\
 & &$N\rho$\hphantom{00}  & \hphantom{0}0.775 &635.11  & 1.5\\
\hline
\end{tabular}
\end{center}
\end{table}

For the nucleon pole $N$ and nucleon resonances $N^*$, the following form factors are used~\cite{xiephi2,Mosel,
feuster1,feuster2}:
\begin{equation}
F_{N}(q^{2}) = \frac{\Lambda_{N} ^{4}}{\Lambda_{N}
^{4}+(q^{2}-m_{N}^{2})^{2}},
\end{equation}
\begin{equation}
F_{N^{*}}(q^{2}) = \frac{\Lambda_{N^*} ^{4}}{\Lambda_{N^*}
^{4}+(q^{2}-M_{N^{*}}^{2})^{2}},
\end{equation}
with $\Lambda_N = 1.0$ GeV and $\Lambda_{N^*} = 2.0$ GeV.

The meson propagators used in our calculation are:
\begin{equation}
G_{\pi / \eta}(k_{\pi / \eta})=\frac{i}{k_{\pi / \eta}^2-m^2_{\pi /
\eta}},
\end{equation}
\begin{equation}
G^{\mu\nu}_{\rho}(k_{\rho})=- i \left(\frac{g^{\mu
\nu}-k^{\mu}_{\rho}k^{\nu}_{\rho}/k^2_{\rho}}{k^2_{\rho}-m^2_{\rho}}\right).
\end{equation}

The propagators of the $N^*$ resonances can be written as
\begin{equation}
G_{N^{*}}(q) = \frac{\not \!q+M_{N^{*}}}{q^{2}-M_{N^{*}}^{2} +
iM_{N^{*}}\Gamma_{N^{*}}},\label{nprop}
\end{equation}
for spin-$\frac{1}{2}$ resonances, and
\begin{equation}
G^{\mu\nu}_{N^*}(q) = \frac{-P_{\mu\nu}(q)}{q^{2}-M_{N^{*}}^{2} +
iM_{N^{*}}\Gamma_{N^{*}}},
\end{equation}
with
\begin{equation}
P_{\mu\nu}(q) = -(\not \!q+M_{N^{*}})\left[g_{\mu\nu}-\frac{1}{3}\gamma_{\mu} \gamma_{\nu}-
\frac{1}{3M_{N^{*}}}(\gamma_{\mu} q_{\nu}-\gamma_{\nu}
q_{\mu})-\frac{2}{3M_{N^{*}}^2}q_\mu q_\nu\right ],
\end{equation}
for spin-$\frac{3}{2}$ resonances.

As usual, in
Eq.~(\ref{nprop}) the following energy-dependent total
width of the $N^*(1535)$ resonance, $\Gamma_{N^*(1535)} (s)$, is employed~\cite{liang},
\begin{equation}
\Gamma_{N^*(1535)} (s) = \Gamma_{N^*(1535)\to N\pi}\frac{\rho_{\pi
N}(s)}{\rho_{\pi N}(M_{N^*(1535)}^2)}+\Gamma_{N^*(1535)\to
N\eta}\frac{\rho_{\eta N}(s)}{\rho_{\eta N}(M_{N^*(1535)}^2)},
\end{equation}
where the two-body phase space
factor $\rho_{\pi(\eta)N}(s)$ is
\begin{equation}
\rho_{\pi(\eta)N}(s)= \frac{2 p^{cm}_{\pi(\eta) N} (s)}{\sqrt{s}} =
\frac{\sqrt{[s-(m_N + m_{\pi(\eta)})^2][s -
(m_N-m_{\pi(\eta)})^2]}}{s}.
\end{equation}

The invariant amplitude for the nucleon pole $N$ or nucleon resonances $N^*$ can be expressed as
\begin{eqnarray}
{\cal M}^{N/N^*} = \sum_{i = \pi,~ \eta,~ \rho} {\cal M}_{i}^{N/N^*}, \label{amp}
\end{eqnarray}
\begin{eqnarray}
{\cal M}_{i}^{N/N^*} = \sum_{j = a,~ b,~c,~d} \eta_{j} {\cal M}_{i,j}^{N/N^*},
\end{eqnarray}
where $\eta_a = \eta_d = 1$ and $\eta_b = \eta_c = -1$.
The explicit expressions of ${\cal M}_{i,j}^{N/N^*}$ can be derived straightforwardly according to the Feynman rules. For example, ${\cal
M}_{\pi,a}^{N^*(1535)}$ can be written as
\begin{eqnarray}
{\cal M}_{\pi,a}^{N^*(1535)} & = &g_{\pi NN} g_{\pi N N^*(1535)}
g_{\eta N N^*(1535)}  F^{N N}_{\pi}(k^2_{\pi}) F^{N^*(1535)
N}_{\pi}(k^2_{\pi}) F_{N^*(1535)}(q^2)\nonumber\\ && \times
G_{\pi}(k_{\pi}) \bar{u}(p_3,s_3) \gamma_5 u(p_2,s_2)  \bar{u}
(p_4,s_4) G_{N^*(1535)}(q) u(p_1,s_1),
\end{eqnarray}
where $s_i~(i=1,2,3,4)$ and $p_i~(i=1,2,3,4)$ represent the spin
projection and 4-momentum of the two initial and two final protons,
respectively.

The $pp$ FSI and $p\eta$ FSI are taken into
account by introducing the following enhancement factors
\begin{eqnarray}
F_{pp}(k_{pp})=\frac{k_{pp}+ i \beta}{k_{pp}- i \alpha}, \label{fsipp}
\end{eqnarray}
\begin{eqnarray}
F_{p\eta}(k_{p\eta})=\frac{1}{1- i k_{p\eta} a}, \label{fsipeta}
\end{eqnarray}
where $k_{pp}$ and $k_{p\eta}$ are the internal momenta of the $pp$ and $p\eta$ subsystems, respectively.
The relevant parameters are: $\alpha = 0.1$ fm$^{-1}$, $\beta = 0.5$ fm$^{-1}$\cite{maeda}, and $a =(0.487+i0.171)$ fm~\cite{feuster1,feuster2}.
 The overall final state interaction is the product of these enhancements~\cite{maeda,bernard}:
\begin{eqnarray}
F_{FSI}=F_{pp}(k_{pp})F_{\eta p}(k_{\eta p_3})F_{\eta p}(k_{\eta p_4}), \label{fsitotal}
\end{eqnarray}
where $p_3$ and $p_4$ denote the two final protons.

With the modular square of the full invariant amplitude
${\cal{|M|}}^2=\sum\limits_{N,N^*}|{\cal M}^{N/N^*}F_{FSI}|^2$, the differential cross section for the $pp \to pp \eta$ reaction can be written as
\begin{eqnarray}
d\sigma (pp\to pp \eta)=\frac{1}{4}\frac{m^2_p}{F} \sum_{s_i}
\sum_{s_f} {\cal{|M|}}^2\frac{m_p d^{3} p_{3}}{E_{3}} \frac{m_p d^{3}
p_4}{E_4} \frac{d^{3} p_5}{2 E_5} \frac{1}{2}\delta^4
(p_{1}+p_{2}-p_{3}-p_{4}-p_5), \label{eqcs}
\end{eqnarray}
where $F$ is the flux factor
\begin{eqnarray}
F=(2 \pi)^5\sqrt{(p_1\cdot p_2)^2-m^4_p}~. \label{eqff}
\end{eqnarray}
The factor $\frac{1}{2}$ before the $\delta$ function in
Eq.~(\ref{eqcs}) comes from the two identical protons in the final
states. The interference terms between different resonances are
ignored.

\section{Numerical results and discussions}

With a Monte Carlo multi-particle phase space integration program,
the total cross section versus excess energy $\varepsilon$ up to 80
MeV, the invariant mass spectra, angular distributions, and Dalitz
plots at excess energies $\varepsilon$ = 15, 40, 57, and 72 MeV for
the $pp \to pp \eta$ reaction are calculated.

\begin{figure}[htbp]
\includegraphics[scale=0.8]{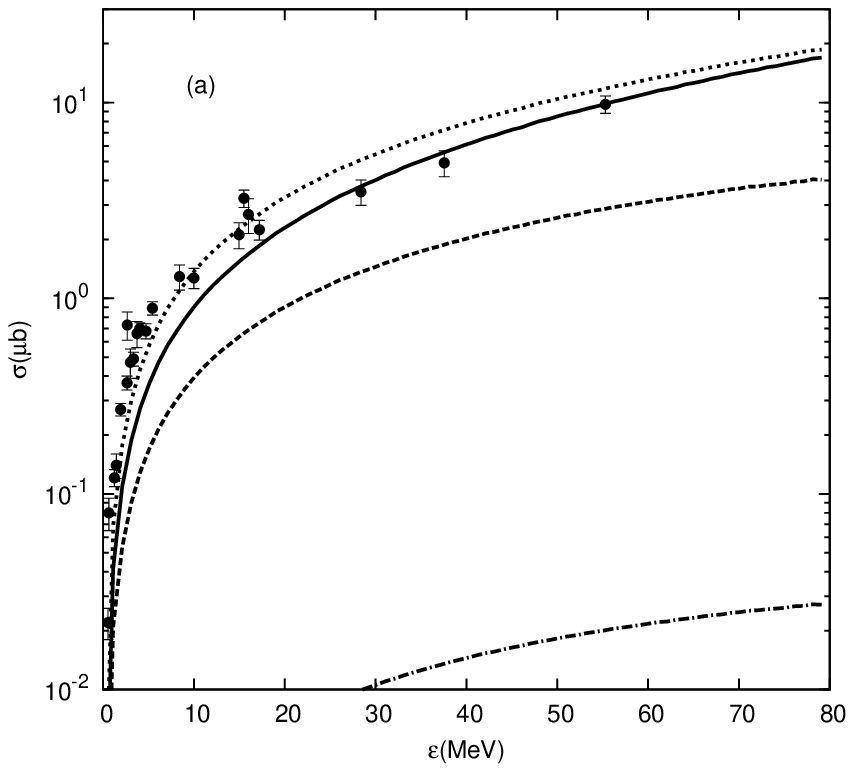}%
\includegraphics[scale=0.8]{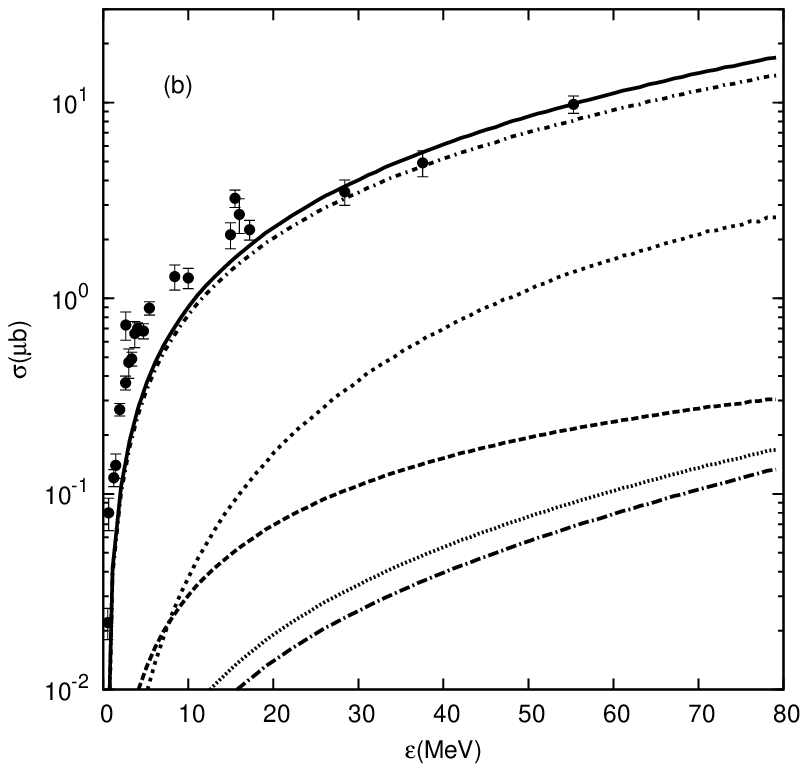}%
\caption{Total cross section vs excess energy $\varepsilon$ for
the $pp \to pp \eta$ reaction from the present calculation (solid
curves) are compared with experimental
data~\cite{eta_exp1,eta_exp2,eta_exp3,eta_exp4,eta_exp5,eta_exp6,calen,balestra,moskal,abdel}.
(a): The dashed, dotted, and dashed-dotted lines stand for
contributions from the $\pi$, $\rho$, and $\eta$-meson exchanges,
respectively. (b): The dashed, dotted, short-dotted, dashed-dotted,
and dot-short-dashed lines stand for contributions from the
$N$, $N^*(1535)$, $N^*(1650)$, $N^*(1710)$, and $N^*(1720)$, respectively.} \label{tcs}%
\end{figure}

The total cross section is shown in Fig.~\ref{tcs} together with the
experimental data. Our results agree fairly well with the experimental
data. From Fig.~\ref{tcs}, one can see that the contributions from
the t-channel $\rho$ and $\pi$-meson exchanges are important and the
$\rho$ exchange plays the dominant role, but the contribution from
the $\eta$-meson exchange is negligible. Fig.~\ref{tcs} also shows
that the contributions from the $N^{*}(1720)$ and $N^{*}(1535)$ are
important and the $N^*(1720)$ plays the dominant role. The
contribution of the $N^*(1535)$ is smaller than that of the
$N^*(1720)$ due to the strong destructive interference among the
exchange mesons, which is similar to the result of
Ref.~\cite{Nakayama:2008tg}. The contributions from the $N^*(1650)$
and $N^*(1710)$ are negligible.

\begin{figure}[htbp]
\includegraphics[scale=0.7]{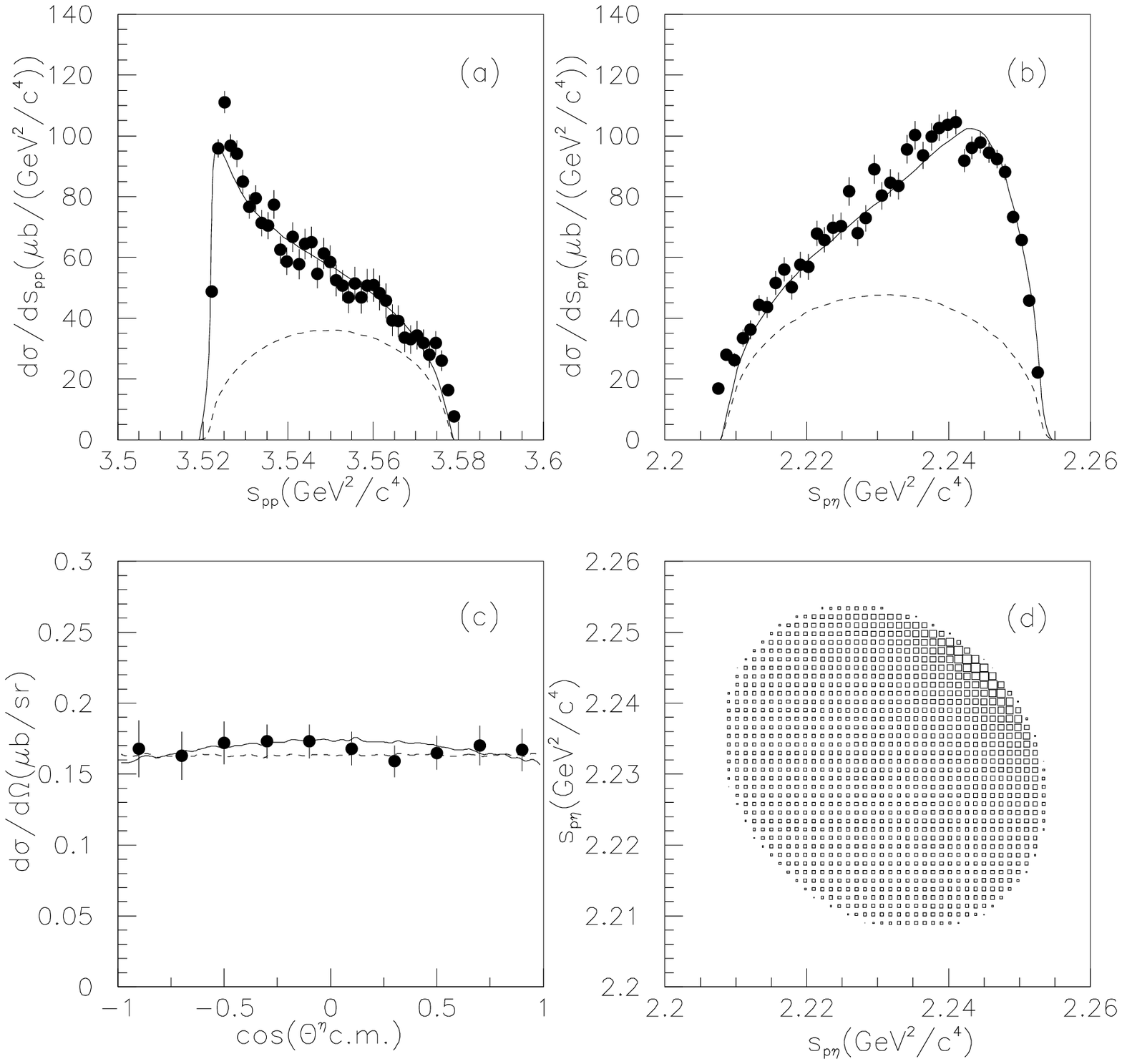}%
\vspace{-0.6cm} \caption{Differential cross sections (solid lines)
and Dalitz plot for the $pp \to pp \eta$ reaction at the excess
energy of $\varepsilon$ = 15 MeV compared with the experimental
data~\cite{moskal,abdel} and phase space distribution (dashed
lines). (a) Distribution of the square of proton-proton
invariant mass; (b) Distribution of the square of
proton-$\eta$ invariant mass; (c) Angular distribution of the
emitted $\eta$ meson in the c.m frame of the total system; (d)
Dalitz plot. }
\label{eta15}%
\end{figure}

The invariant mass spectra, angular distribution, and Dalitz plot
at excess energy $\varepsilon$ = 15 MeV are shown in
Fig.~\ref{eta15} together with experimental data. The measured $pp$
and $p\eta$ invariant mass spectra and the angular distribution of
$\eta$ are reproduced well. From Fig.~\ref{eta15} (a) and (d),
one can see that the $pp$ FSI plays an important role.

\begin{figure}[htbp]
\includegraphics[scale=0.7]{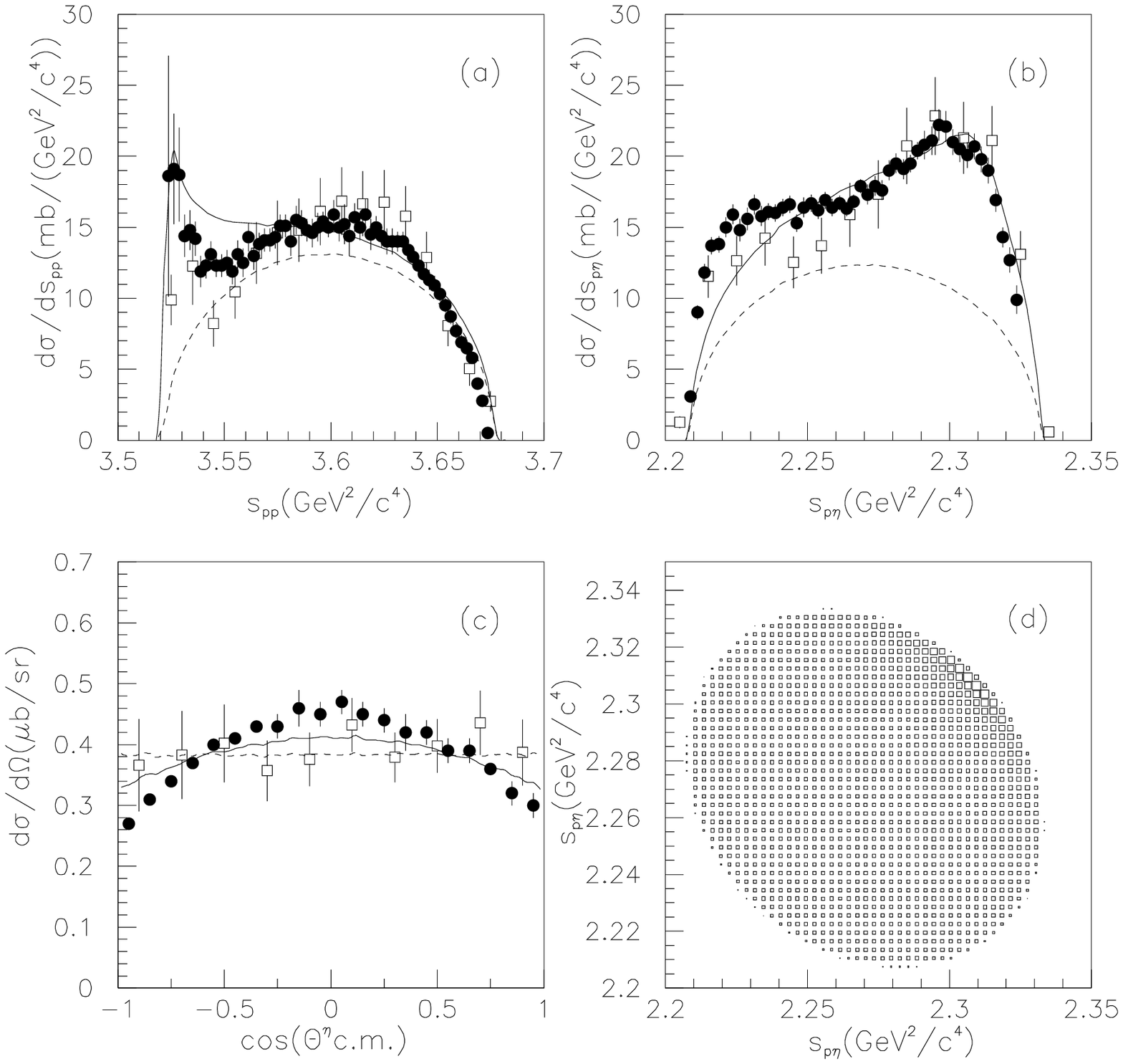}%
\vspace{-0.6cm} \caption{Same as Fig.~\ref{eta15} but at excess
energy of $\varepsilon$ = 40 MeV. Experimental data are taken from
Ref.~\cite{abdel} (squares) and Ref.~\cite{petr} (dots).}
\label{eta41}%
\end{figure}

The invariant mass spectra, angular distribution, and Dalitz plot
at excess energy $\varepsilon$ = 40 MeV as well as the experimental
data are shown in Fig.~\ref{eta41}. For the invariant mass spectra
of proton-proton and proton-$\eta$, the theoretical results are in
agreement with the experimental data except for those near the proton-proton (proton-$\eta$) threshold. This small discrepancy indicates
the $pp$ FSI used in our calculation may be somewhat strong in this
region.

For the angular distribution of the emitted $\eta$ meson in the
overall c.m. frame, there are two groups of data which do not agree
with each other~\cite{abdel,petr}. One is isotropic~\cite{abdel},
while the other is anisotropic~\cite{petr}, as shown in
Fig.~\ref{eta41} (c). Our result indicates that the angular
distribution of the $\eta$ meson is anisotropic, consistent with the
data from Ref.~\cite{petr}. As pointed out by
Ref.~\cite{calen,germond,petr}, the anisotropy is probably due to a
mainly destructive interference between the dominant $\rho$ exchange
and $\pi$ exchange. It is interesting to point out that the
$N^*(1535)$ dominant
interpretations~\cite{shyam07,knakayama,Fix:2003gs} give almost
isotropic angular distribution of the $\eta$ at this region except
that Ref.~\cite{nakayama} gives the anisotropic angular
distribution of the $\eta$ by allowing for contributions from
baryonic and mesonic currents.

\begin{figure}[htbp]
\includegraphics[scale=0.7]{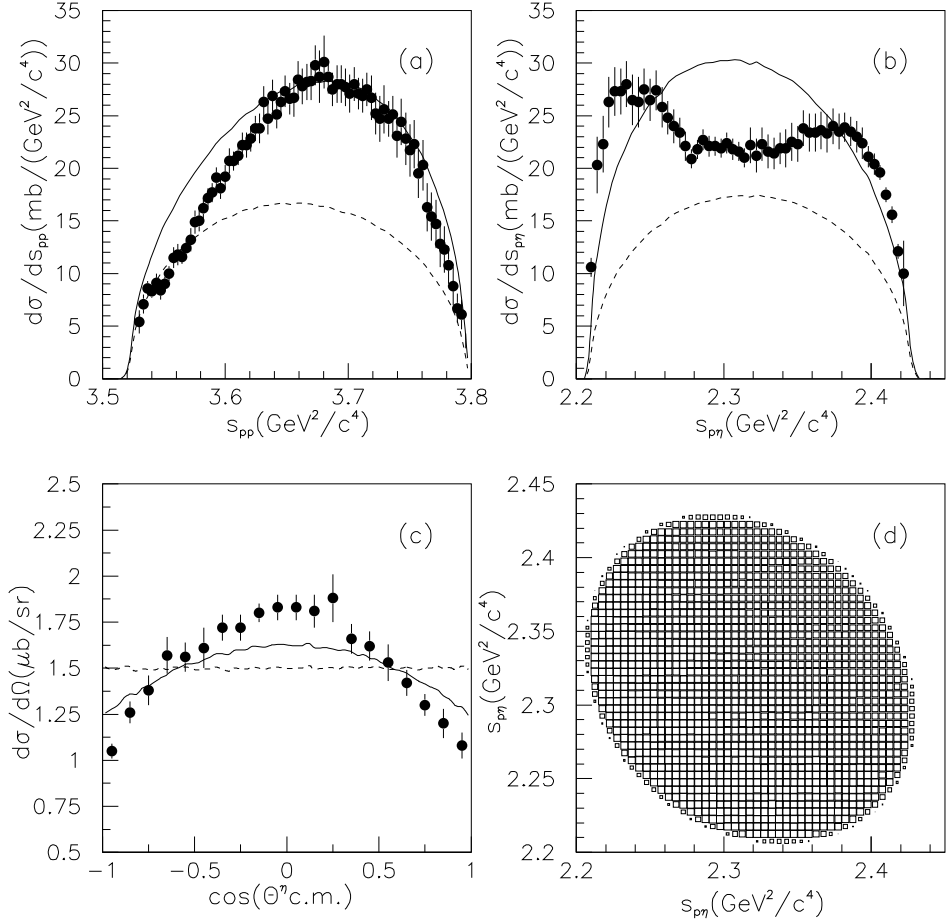}%
\vspace{-0.6cm} \caption{Same as Fig.~\ref{eta15} but at excess
energy of $\varepsilon$ = 72 MeV, and with $pp$ FSI ignored. Experimental data are taken from
Ref.~\cite{petr}.}
\label{eta72}%
\end{figure}

The invariant mass spectra, angular distribution, and Dalitz plot
at excess energy $\varepsilon$ = 72 MeV as well as the experimental
data are shown in Fig.~\ref{eta72}. The experimental data shown in
Fig.~\ref{eta72} (a) indicate the $pp$ FSI should be rather weak, so
the $pp$ FSI is ignored in this energy region. This rough procedure
has been used in double-pion production in nucleon-nucleon collisions
 and the results turn out to be considerably improved~\cite{xuc}.
 Our $pp$ invariant mass spectrum can reasonably account for the
 data.

 The two-peak structure in the proton-$\eta$ distribution cannot be
 reproduced in our calculation, which is similar to the result from Ref.~\cite{Nakayama:2008tg}. This suggests
  that the structure in the $p\eta$ distribution
 cannot be simply interpreted by the
$N^*(1535)$, $N^*(1650)$, $N^*(1710)$, and $N^*(1720)$ resonances,
and a more complicated mechanism is strongly called for.

Our angular distribution of the $\eta$ at $\varepsilon$ = 72 MeV again
indicates that the $\eta$ distribution is anisotropic, consistent
with the data from Ref.~\cite{petr}. To our knowledge, there is as
yet no theoretical paper for addressing the angular distribution of
the $\eta$ at this region.

\begin{figure}[htbp]
\includegraphics[scale=0.7]{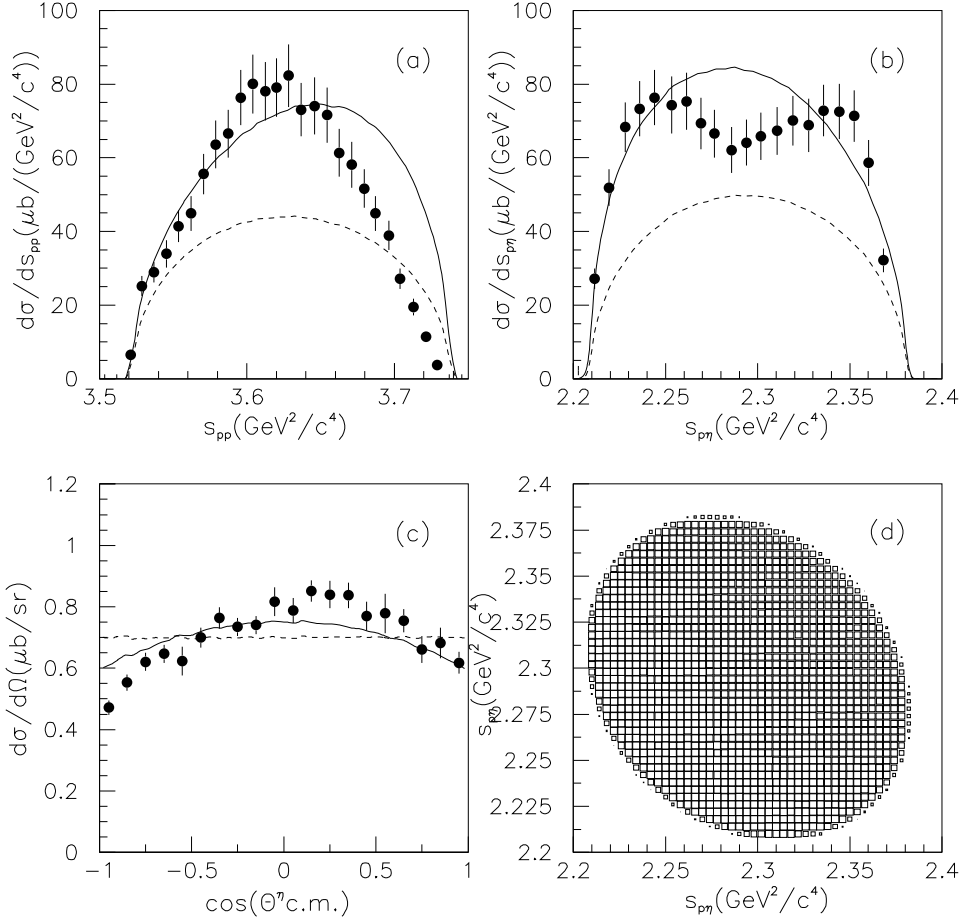}%
\vspace{-0.6cm} \caption{Same as Fig.~\ref{eta15} but at excess
energy of $\varepsilon$ = 57 MeV, and with $pp$ FSI ignored. Experimental data are taken from
Ref.~\cite{Shah:2011zz}.}
\label{eta57}%
\end{figure}

The invariant mass spectra, angular distribution, and Dalitz plot
at excess energy $\varepsilon$ = 57 MeV as well as preliminary experimental
data are shown in Fig.~\ref{eta57}. Similar to the case at excess energy $\varepsilon$ = 72 MeV, the preliminary data of Ref.~\cite{Shah:2011zz} show that the two-peak structure appears in the proton-$\eta$ distribution and the angular distribution of the $\eta$ is anisotropic. Our predicted angular distribution for the $\eta$ at $\varepsilon$ = 57 MeV is anisotropic, consistent with the data. However,
the $p\eta$ invariant mass distribution shows large differences between the present model and the experimental data.

It is noted that our present model does not include the higher partial waves for the $p\eta$ FSI. As pointed out by Refs.~\cite{petr,Krusche:2014ava}, the experimental data indicate the higher partial waves at higher reaction energies could be important. The contributions of the higher partial waves being neglected in the current model calculations may cause the discrepancy between the experimental data and our model of the proton-$\eta$ distribution at excess energies $\varepsilon$ = 57 and 72 MeV.

\section{Summary and Conclusions }

In this paper we have calculated the $pp \to pp \eta$ reaction within an
effective Lagrangian approach combined with the isobar model. Our model calculations
can reasonably reproduce the total cross sections up to excess
energy of 80 MeV.

It is shown that for the $pp \to pp\eta$ reaction, the contribution
of the $\rho$-meson exchange is larger than that of the $\pi$-meson
exchange, and the contribution of the $N^*(1720)$ is larger than
that of the $N^*(1535)$.

Also, the same cut-off parameters for the $N^*(1650)$, $N^*(1710)$,
and $N^*(1720)$ resonances are used, which makes it suitable to
investigate the relative contributions of the $N^*(1650)$,
$N^*(1710)$, and $N^*(1720)$ resonances.  Our results show that the
contributions from the $N^*(1650)$ and $N^*(1710)$ are negligible.

Our calculations can reasonably explain the measured $pp$ and
$p\eta$ invariant mass spectra at excess energies $\varepsilon=$ 15 and 40 MeV, but
fail to explain the two-peak structure in the proton-$\eta$
distribution at excess energies $\varepsilon=$ 57 and 72 MeV, which suggests that in the higher energy
region, a more complicated mechanism is needed.

We give the anisotropic angular distribution of the $\eta$ at
$\varepsilon$ = 40, 57 and 72 MeV, consistent with the data from
Refs.~\cite{petr,Shah:2011zz}. This favors the interpretation that the
interference between the $\rho$ exchange and $\pi$ exchange is
mainly destructive.

\bigskip
\noindent
{\bf Acknowledgement}\\

We thank Dr. Ju-Jun Xie for helpful discussions and valuable
suggestions. This work is partly supported by the National Natural
Science Foundation of China under grant 11105126.

\end{document}